\documentclass{iaus}
\usepackage{graphicx}

\title[Microlensed dwarf stars in the Bulge] 
{Elemental abundances in the Galactic bulge from microlensed dwarf stars}

\author[T. Bensby et al.]   
{T. Bensby,$^1$
 S. Feltzing,$^2$
 J.A. Johnson,$^3$
 A. Gould,$^3$
 H. Sana,$^4$\\
 A. Gal-Yam$^5$
 M. Asplund$^6$
 S. Lucatello$^7$
 J. Melendez$^8$\\
 A. Udalski$^9$
 D. Kubas$^{10}$
 G. James$^{11}$
 D. Ad\'en$^2$
 \and
 J. Simmerer$^2$}
\affiliation{$^1$European Southern Observatory, Santiago, Chile  \\[\affilskip]
$^2$Lund Observatory, Lund, Sweden  \\[\affilskip]
$^3$Dept of Astronomy, Ohio State University, Columbus, Ohio, USA \\[\affilskip]
$^4$Univ. van Amsterdam, Sterrenkundig Instituut 'Anton Pannekoek',
Amsterdam, Netherlands \\[\affilskip]
$^5$Benoziyo Center for Astrophyics, Weizmann Institute of Science, Rehovot, Israel \\[\affilskip]
$^6$Max Planck Institute for Astrophysik, Garching, Germany \\[\affilskip]
$^7$INAF-Astronomical Observatory of Padova, Padova, Italy \\[\affilskip]
$^8$Centro de Astrofisica da Universidade do Porto, Porto, Portugal \\[\affilskip]
$^9$Warsaw University Observatory, Warszawa, Poland \\[\affilskip]
$^{10}$Institut d'Astrophysique de Paris, Paris, France \\[\affilskip]
$^{11}$European Southern Observatory, Garching, Germany}
\pubyear{2009}
\volume{265}  
\pagerange{XX--XX}
\jname{Chemical abundances in the Universe -- Connecting first stars to planets}
\editors{Katia Cunha, Monique Spite \& Beatrice Barbuy, eds.}
\begin{document}

\maketitle

\begin{abstract} 
We present elemental abundances of 13 microlensed dwarf and subgiant stars in the Galactic bulge, which constitute the largest sample to date. We show that these stars span the full range of metallicity from $\rm Fe/H=-0.8$ to $+0.4$, and that they follow well-defined abundance trends, coincident with those of the Galactic thick disc. 
\end{abstract}



The formation and evolution of bulges are an integral and central 
aspect of galaxy formation and evolution. Much of what we know about 
the formation and evolution of the Milky Way bulge comes from giant 
stars.  However, the underlying  assumption that the giants 
accurately represent all the stars has not yet been rigorously 
tested (e.g., Santos et al.~2009). A true picture of the star 
formation history in the Galactic bulge requires the study of dwarf 
stars. Under normal circumstances, at the distance of the Galactic 
bulge, dwarf stars are too faint to acquire the high-resolution 
spectra that are crucial for abundance analysis. However, 
microlensing events offer, during a short time, the opportunity 
to obtain spectra of dwarf stars in the Bulge.

In 2009 we have obtained high-resolution spectra for several microlensed 
dwarf and subgiant stars using UVES on the VLT.  Combined with previous 
events (Johnson et al. 2007, 2008; Cohen et al. 2008, 2009; 
Bensby et al. 2009a, 2009b) the total number of  dwarf and sub-giant 
star in the Bulge that we have analysed is now 13. 

\begin{figure}
\resizebox{\hsize}{!}{
\includegraphics[bb=-190 160 800 450,clip]{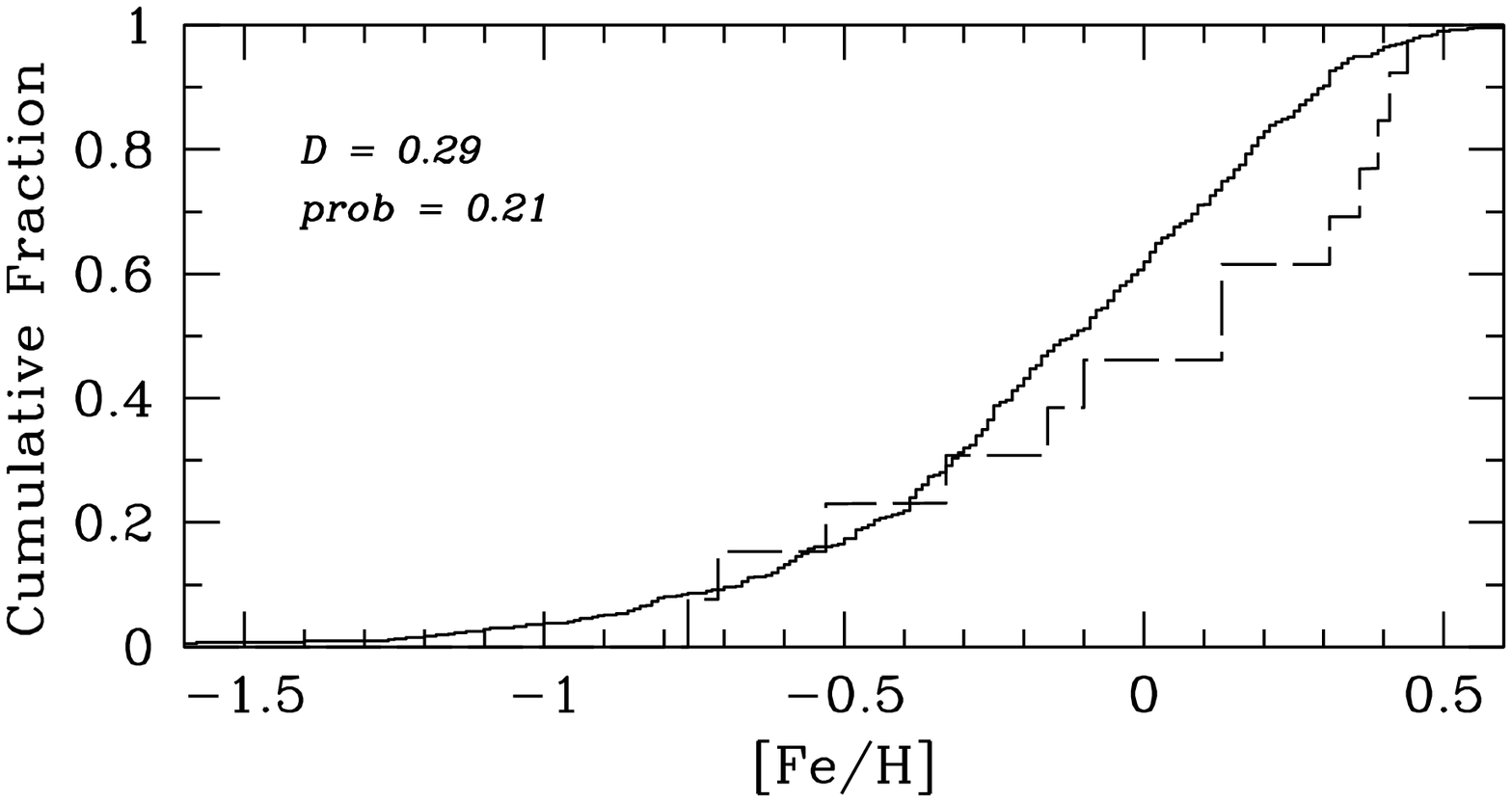}}
\resizebox{\hsize}{!}{
       \includegraphics[bb=0 165 610 500,clip]{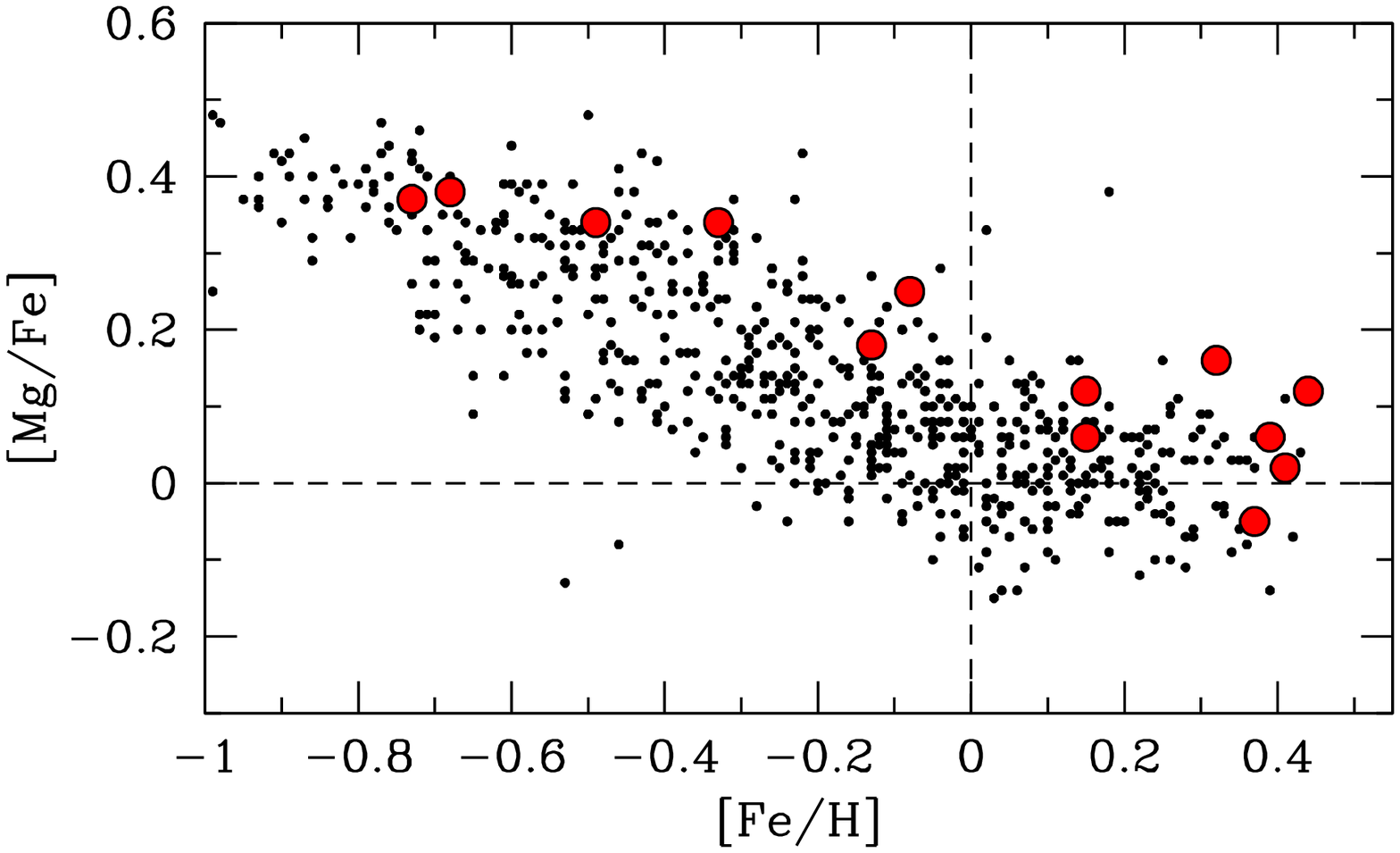}
       \includegraphics[bb=0 165 610 500,clip]{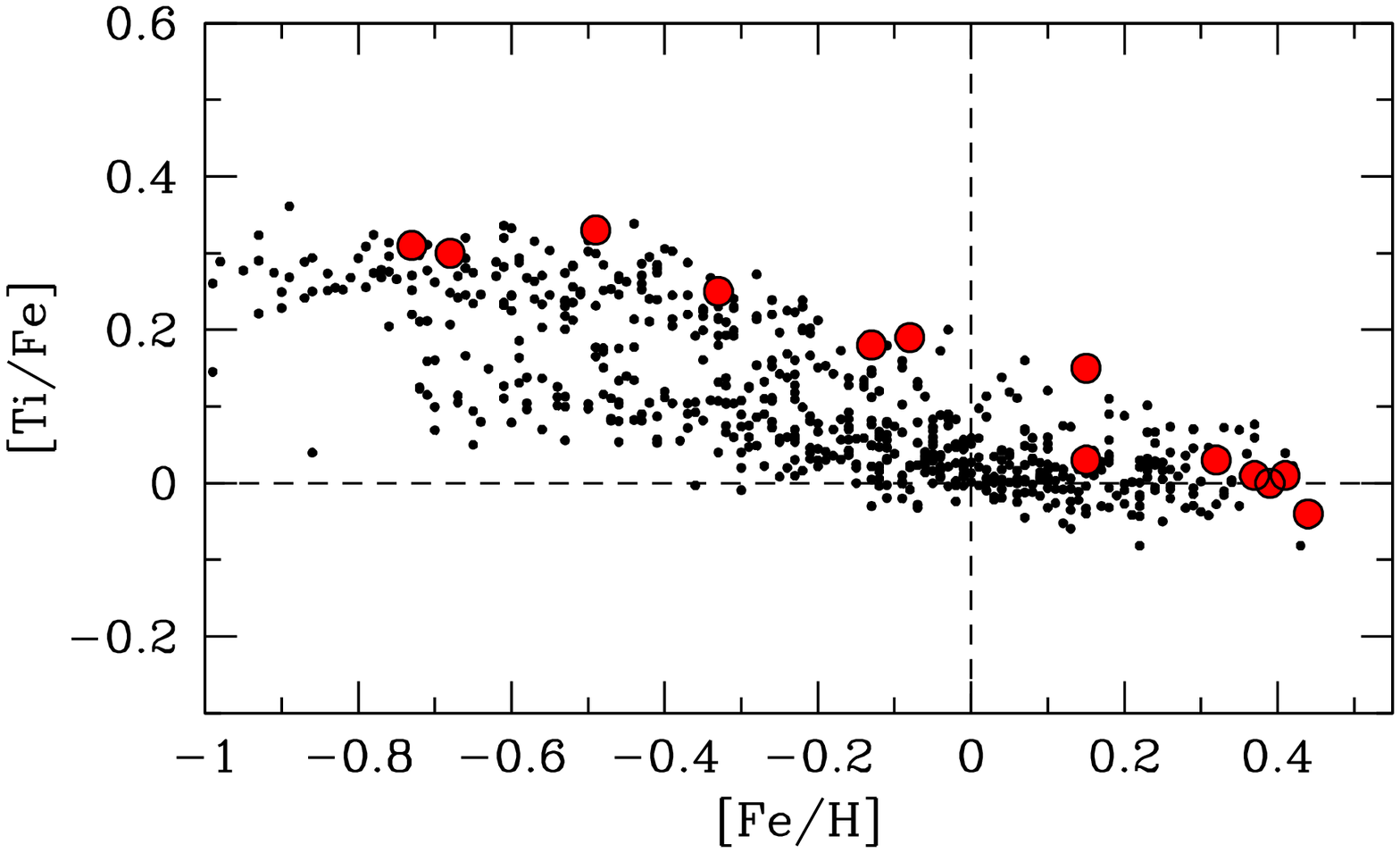}}
\caption{{\it Top panel} shows the two-sample KS-test between 
the Zoccali et al.~(2008) giant stars (full line) and 
the 13 microlensed dwarf and subgiant stars (dashed line).  
{\it Bottom panels} show        [X/Fe] versus [Fe/H]. Thin and thick 
        disc stars from Bensby et al.~(2003, 2005, and 2010 in prep)
        are marked by black dots, and the 
        microlensed Bulge dwarfs by red (larger) circles.
         \label{fig:kstest}}
\end{figure}

These 13 microlensed dwarf and subgiant stars have an 
average metallicity of $\rm \langle [Fe/H]\rangle=-0.03\,\pm0.4$.   
A two-sided KS-test (see top panel in Fig.~\ref{fig:kstest}) does 
not allow us to 
reject the null-hypothesis that the MDF from the  microlensed dwarfs 
and the MDF for the 500 Bulge giants from  Zoccali et al.~(2008) are 
identical, even  adopting a loose significance level of  0.1. 
More microlensed events would help refining the comparison.
It is, however, evident that the super-metal-rich MDF proposed by 
Cohen et al.~(2009), is starting to shift toward lower metallicities.

Bottom panels of Fig.~\ref{fig:kstest}  show the abundance 
trends for the microlensed Bulge dwarf and sub-dwarf stars compared 
to nearby thin and thick disc dwarf stars. Two things can be taken 
from this figure: 1) The bulge stars do have a wide spread in metallicity; 
2) The microlensed Bulge dwarfs fall together on the plot with dwarf
stars that are known to belong to the Galactic thick disc, i.e., 
high alpha-element abundances relative to iron. 

Regarding the Bulge membership for these microlensed dwarf stars, 
theoretical calculations for the distance to microlensed sources, 
assuming a constant disk density and an exponential bulge, show that 
the distance to the sources is strongly peaked in the Bulge, with 
the probability of having $D< 7$ kpc very small (Kane \& Sahu~2006). 
Also, with regard to the sources being in the disk on the other side 
of the Bulge, the position of the stars in the OGLE/MOA 
color-magnitude diagram strongly argues against 
that. Basically, the stars are not faint enough, especially when 
considering the fact that there will be additional dust as we enter 
the disc on the other side. 

The data set, the analysis, and results for other $\alpha$-elements 
will be presented in Bensby et al.~(in prep.),
and results for $r$- and $s$-process elements in Johnson et al.~(in prep.).



\begin{thebibliography}{}


\bibitem[]{bensby2009letter}{Bensby}, T., {Feltzing}, S., {Johnson}, J.~A., {et~al.} 2009b,  {\it ApJ}, 699, L174

\bibitem[]{bensby2003}
{Bensby}, T., {Feltzing}, S., \& {Lundstr{\" o}m}, I. 2003, {\it A\&A}, 410, 527

\bibitem[]{bensby2005}{Bensby}, T., {Feltzing}, S., {Lundstr{\" o}m}, I., \& {Ilyin}, I. 2005, {\it A\&A},  433, 185

\bibitem[]{bensby2009}
{Bensby}, T., {Johnson}, J.~A., {Cohen}, J., {et~al.} 2009a, {\it A\&A},
  499, 737
 
\bibitem[]{cohen2008}{Cohen}, J.~G., {Huang}, W., {Udalski}, A., {Gould}, A., \& {Johnson}, J.~A.
  2008, {\it ApJ}, 682, 1029

\bibitem[]{cohen2009}
{Cohen}, J.~G., {Thompson}, I.~B., {Sumi}, T., {et~al.} 2009, {\it ApJ}

\bibitem[]{johnson2007}
{Johnson}, J.~A., {Gal-Yam}, A., {Leonard}, D.~C., {et~al.} 2007, {\it ApJ}, 655,
  L33

\bibitem[]{johnson2008}
{Johnson}, J.~A., {Gaudi}, B.~S., {Sumi}, T., {Bond}, I.~A., \& {Gould}, A.
  2008, {\it ApJ}, 685, 508

\bibitem[]{kane2006}
{Kane}, S.~R., \& Sahu, K.~C., 2006, {\it ApJ}, 637, 752

\bibitem[]{santos2009} 
{Santos}, N.~C., {Lovis}, C., {Pace}, G., {Melendez}, J., \& {Naef}, D., 2009, {\it A\&A}, 493, 309

\bibitem[]{zoccali2008}
{Zoccali}, M., {Hill}, V., {Lecureur}, A., {et~al.} 2008, {\it A\&A}, 486, 177

\end{thebibliography}
\end{document}